# Translating cities
## the use of digital technologies in urban environments
Working Paper

**Gloria Lanci**
University of the West of England – UK
gloria.lanci@uwe.ac.uk

**Abstract**

Computer models and information systems have been used for urban planning and design since the 1950s. Their capacity for analysis and problem-solving has increased substantially since then with hardware and software being able to manage large amounts of data. The beginning of the 2000s brought better technologies for data visualisation and intuitive software products and nowadays they are being used to design and manipulate highly complex urban systems.

However, ontological and epistemological questions remain about the nature of urban environments. What do we know about cities - and what is a city exactly? What theoretical models can be applied to the study of cities? How do we translate a city into data? What type of information really matters and what do we want to communicate with them? What are the implications of computer modelling for urban planning and design?

This paper reviews how digital technologies have been used to shape our understanding of cities – implicating the development of urban theories – and how they impact the design and planning of cities. The point of depart is a scrutiny of the emergence of modern planning in the nineteenth century when cities became a 'scientific subject'. A range of theories and concepts of the urban form and urban growth developed during the twentieth century will be presented and then linked to the first investigations involving the use of the computer for modelling and planning. The paper concludes with some discussions around the interaction of computing and urban theories.

1. **The emergence of modern urban planning**

The nineteenth century was a period when urbanisation reached unprecedented levels of growth. Overcrowded dwellings, disease, pollution and noise reached new levels, a collective nightmare that urged solutions at the large-scale. In the earlier years of the modern urban planning two main objectives were settled: to *sanitise* the city, turning it into a healthier place to live, and to *reform* the city living itself, providing modern and more efficient public urban infrastructure combined with better dwellings. These objectives were addressed through physical improvements and social programmes that in many senses aimed to build a new urban society (Cherry 1996; Hall 1988).

The state, the church and professionals, including architects, engineers and doctors, were compelled to think about urban settlements in a systematic way. Cities were translated into numbers and codes; surveys and statistics of all sorts were produced. In the UK for instance regular national censuses started to take place as early as in 1801; detailed accounts of the social condition of the working class were conducted throughout the Victorian period[1] and the investigation of John Snow on the cholera outbreak in the centre of London in 1854 is now considered one of the first applications of a geographical information system (Heywood 2011; Maguire 2005).

The conceptual framework developed during the nineteenth century in relation to urban planning could be described as a systematic approach involving consecutive phases: defining a problem, collecting data, analysing data and providing solutions. The resulting city plan would then be presented to the authorities and general public in form of schematic drawings, maps and elaborated perspectives of the future landscape.

In Europe the plan to modernise and extend Barcelona conceived by the Spanish engineer Idelfons Cerda (1815-1876) in the 1860s is a good example of a large-scale project developed under this framework. Cerda started his career conducting topographical surveys and statistical studies for several public offices. His proposal, known as Plan Cerda, was embedded with ideas of order, categorisation and hierarchy. Cerda is often considered the founding father of modern urban planning and credited as having coined the term "urbanisation".

The propagation of the concepts and methods of re-creating cities soon lead to criticism. One of the most influential works is probably *City planning according to artistic principles* written by Camillo Sitte (1843-1903) and published in 1889. Sitte was an Austrian architect and art historian who saw modern planning as too pragmatic and hygienic, dismissing the symbolic and aesthetic values of ancient (meaning medieval) urban environments. Sitte praised irregularity, monumentality and beauty as components of good design.

Resonances of these ideas can be found in the Garden City and City Beautiful movements. The former was developed in Britain and was mostly a creation of the planner and entrepreneur Ebenezer Howard (1850-1928) and was described in his book *Garden Cities of To-morrow*, published in 1898; the latter flourished in North America during 1890s and 1900s and had the architect Daniel Burnham (1846-1912) as one of its main advocates.

Exploring theories from diverse disciplines Patrick Gueddes (1854-1932), a Scotish biologist, sociologist, geographer and planner, introduced innovative approaches in the field of urban planning. He adopted biological concepts of evolution in social analysis and considered the human interaction with its surrounding natural environment as crucial for urban planning, extending geographical survey to the scale of the region where cities were located. His theories were presented in his book *Cities in Evolution*, published in 1915.

## 2. Theories of urban design and urban form

Urban and regional planning was institutionalised in most of the European and American cities in the first half of the twentieth century. This period also saw the emergence of the modern architecture movement that had a significant influence in urban planning through the *Congrès Internationaux d'Architecture Moderne* (CIAM), the International Congress of Modern Architecture, which lasted from 1928 to 1959.

The CIAM was the organization that spread the most up-to-date ideas of its time in urbanism. One of its founding members Le Corbusier (1887-1965) saw the house as "a machine to living in" and the city a product of functional ordered spaces, which would provide efficient solutions for housing, working and leisure (Hall 1988).

---

[1] Such as the classic study of Engels, *Die Lage der arbeitenden Klasse in England* (The condition of working class in England), written during his stay in Manchester between 1842 and 1844, and the extensive survey undertaken by Charles Booth between 1886 and 1903, published in a series of volumes entitled *Life and Labour of the People in London*.

After the Second World War cities in Europe faced the challenge of reconstruction and with that came another wave of proposals in planning and design, combined with mass production of housing dwellings more or less inspired by the modern architecture movement. However by the late 1950s this model of "efficiency", "simplicity" and "functionality" started to receive criticism from both the academic/expert side as well as from the lay public, especially concerning housing developments.

*Urban design*

A new theoretical *corpus* was developed around the concept of urban design. It incorporated an interdisciplinary approach to urban planning and architecture that demanded a "return to origins within the built environment professions" through "classic principles" such as walkable streets, human-scale buildings, and an active public realm (Larice and Macdonald 2007).

Most of the current theory in urban design is founded upon what Marshal (2012) called the "four classic urban design treatises": *The image of the city* (Lynch, 1960), *Townscape* (Cullen, 1961), *The death and life of great American cities* (Jacobs, 1961) and *A city is not a tree* (Alexander, 1965). They are frequently cited and widely accepted as fundamentals for "good" principles of design, although they were not able to establish a scientific approach of the discipline since the methods proposed were not systematically tested and their "original hypotheses remain largely unconfirmed" (Marshall, 2012). These works have in common a critique of an artificial or non-human approach to design that could be translated as a rejection of the modernist experiences in Europe and United States post Second World War.

Kevin Lynch (1918-1984) sought to find empirical evidence that people hold a series of images related to the urban environment and that these images could be categorised into five basic elements: landmarks, nodes, districts, edges and paths. Lynch suggests that through identifying and analysing these elements urban designers would have a guideline for better practice (Lynch 1960).

Gordon Cullen's (1914-1994) main work, *Townscape*, first published in 1961, is a study of the sensorial human experience on urban landscapes. He mostly explores the visual impact that different urban environments have upon us and introduces the concept of "serial vision" or the sequence of scenarios that one would see while walking through the urban environment. Cullen proposed that "good" urban places are those able to give "visual pleasure" and urban designers should try to exercise the "art of relationship" combining elements that would achieve this (Cullen 1961).

Jane Jacobs (1916-2006) was a journalist and writer whose critique of urban planning and design turned into a manifesto for better designed public spaces. She is regarded as the inspirational source for the foundation of the New Urbanism movement in the United States in the 1980s, which promotes car-free, walkable neighbourhoods and more diverse and humanised cities. Jacobs stated that four physical conditions should be observed by planners for a dynamic urban life: multifunctional neighbourhoods, short blocks and connected street networks, varied age residential areas and high density of population. She contended that these conditions were necessary to keep cities "as living organisms" in which streets were the "lifeblood" (Jacobs 1961).

Christopher Alexander's (1936-) most well-known article *A city is not a tree* states that "natural cities", those that were developed more or less spontaneously, have a semi-lattice structure and this is what makes them complex and diverse spaces. On the other hand the "artificial cities", or those that were created by planners and designers, have a tree structure that results in monotonous apathetic spaces. Alexander advocates the semi-lattice structure as the model to be followed by designers in order to create a living city (Alexander, 1965). His argument is based on mathematical concepts and it is illustrated with diagrams that explain the connections between elements in a set or network. In 1977 he published *A pattern language* where he presents a method for good design practice, setting general rules that resembles a generative grammar[2]. Again Alexander is seeking to identify the features which make "natural cities" so attractive and pleasant to live in, looking to their elements, structure and compositions (Alexander, 1965).

*Urban morphology*

Urban morphology was developed by a group of theoreticians who were interested in describing and analysing urban environments by "decomposing" them into categories, elements or types in order to understand what constitutes cities.

The two central figures of this theory are M. R. G. Conzen (1907 – 2000), a German geographer who settled as practitioner and lecturer in England, and Salvatori Muratoni (1910-1973), an Italian architect who was a lecturer in Venice and Rome. They started the English and Italian schools of urban morphology in the early 1960s and were joined by French architects Philippe Panerai (1940-) and Jean Castex (1942-) in the late 1960s (Moudon 1997; Whitehand 1992).

These three schools – the Conzenian, Muratorian and Versailles – were only brought together in 1994 with the establishment of the ISUF, the International Seminar on Urban Form. By that time researchers in urban morphology were working in US, Germany, Spain, Poland, Austria and Portugal, and the Versailles school had stretched its influence into the Latin and Arab worlds (Moudon 1997).

The theoretical basis of urban morphology relies on the statement that cities can be "read" and analysed through their physical form and that morphological analysis must involve three fundamental components: *form*, *resolution* and *time* (Moudon 1997). Urban *form* is defined by buildings and their related open spaces such as plots and streets; the *resolution* is related to the scale of the urban form and can have four gradients corresponding to the building/lot, the street/block, the city and the region; *time* is the historical perspective under which transformations in urban form can be understood. Conzen proposed several key elements of the urban form; the most relevant ones being land use, building structure, plot pattern and street pattern (Moudon 1997).

## 3. Understanding cities in the computer age

The 1950s and 1960s brought computers and information systems into architecture and urban planning practices. This period coincides with the quantitative revolution in geography and the increased use of statistical techniques and mathematical models in research (Holton-Jensen 1999).

The advances in computing technology throughout the second half of the twentieth century made it possible to work with more complex models and to conduct more extensive analysis. Previous conceptual models of urban growth such Weber's *Industrial Location Model* (1909), Burgess' *Concentric Zone Model* (1925) and Christaller's *Central Place Theory* (1933) although very influential were static in nature, disregarding temporal aspects of urban development (Liu 2009). These and subsequent theories inspired by them treated cities as closed systems in constant equilibrium and not affected by the wider environment (Batty 2012).

---

[2] According to the Oxford Dictionary *generative grammar* is "a type of grammar which describes a language in terms of a set of logical rules formulated so as to be capable of generating the infinite number of possible sentences of that language and providing them with the correct structural description".

The development of Geographical Information Systems (GIS) from the 1960s onwards made analysis methods, such as McHarg's map overlaying, more practical and decision making more intuitive (Goodchild 1992; Goodchild and Hening 2004). Progress in information systems in the 1980s and 1990s had significant impact for urban design and urban modelling (Batty et al 1998; Liu 2009). Visualisation improved with new platforms made possible the transfer of Computer Aided Design (CAD) data into GIS, allowing the development of numerous 3-D models (Batty et al 1998). Towards the end of 1990s GIS had developed into a user-friendly environment able to link spatial attributes and quantitative data at a fine scale and to produce sophisticated spatial analysis and maps (Batty et al 1998; Moudon 1997). The city was incorporated in the digital world in full.

*Space syntax*

Space syntax was first conceived by a group of researchers at University College London during the late 1970s. Its concepts and techniques were presented in the book *The social logic of space*, (Hillier and Hanson 1984). Their model is an application of the theory of graphs to the analysis of spatial configuration of urban spaces, connecting them with their level of usage (Jiang, Claramunt and Klarqvist 2000).

The theoretical approach of space syntax is to consider relations between spaces and their potential to "embody or transmit social ideas", providing a measurable scale from segregation to integration and a platform to analyse the social level of interaction in a specific space (Hillier and Hanson 1984; Hillier and Vaughan, 2007). Space syntax establishes therefore a sort of "social geometry": human movement is essential linear (corridors, streets) while social interaction takes place in convex spaces (rooms, squares) and in any configuration it is possible to determine visual fields or isovists[3] (Hillier and Vaugha, 2007).

In space syntax the urban environment is modelled at the cognitive level. For instance, open spaces as streets are represented according to their "lines of sight" or axial lines and the nodes in the urban system are where these lines intersect. This network of lines and nodes can be then measured and categorised as less or more integrated in the urban system (Batty 1998).

The concepts of space syntax are: the human perception of space or 'scale space vision'; the spatial decomposition that allows spaces to be categorised into the three main types (linear spaces, convex spaces and isovists; and the spatial analysis parameters for connectivity, control value and integration (Jiang, Claramunt and Klarqvist 2000).

Since 1989 space syntax has been implemented by a company offering consulting services to governments, organisations and private stakeholders. Its methodology has been continuously developed and is employed widely in planning, design, transport and property development through this consultancy.

Space syntax has, however, been criticised for its inconsistencies in representing urban spaces (Ratti 2004) and praised for offering inspired alternatives for spatial configuration analysis such as the Urban Network Analysis (UNA) toolbox (Sevtsuk and Mekonnen 2012).

---

[3] Isovists were introduced by Benedikt in 1979.

*Complex systems*

> Why have cities not, long since, been identified, understood and treated as problems of organized complexity? (Jacobs 1961)

The complex nature of cities has been long ago recognised[4] but its replication in modern urban planning has not reached the diversity in function and structure of traditional cities. This remains the core challenge for planners and designers according to Jane Jacobs and apparently it has been 'haunting' researchers in the digital age.

Michael Batty, an urban planner and geographer, has been combining computer sciences with urban research through his career since 1970s, following the main developments in both fields. After 1990s and in collaboration with other researchers he has explored fractal, chaos and complexity theories.

Fractal geometry or the "theory of roughness" was first proposed by Benoit Mandelbrot in 1975; his most well-known work *The Fractal Geometry of Nature* was published in 1982[5]. One of the first attempts to apply fractal geometry to the study of cities is presented in the book *Fractals Cities*, written by Michael Batty and Paul Longley. In the introduction the authors state that their main argument relies on the fact that cities are fractal in form and that "much of our pre-existing urban theory is a theory of the fractal city" (Batty and Longley 1994).

The study of complex systems can be found in many disciplines from anthropology to neurosciences and it often involves the development of mathematical models (Liu 2009; Portugali 2012). Batty currently pursues the establishment of a "science of cities" that is much based on new approaches derived from complexity sciences, making the transition from thinking of "cities as machines" to "cities as organisms" (Batty 2012).

Nikos Salingaros, a mathematician and urban theorist who since 1983 has been collaborating with Christopher Alexander, pointed to the need for a scientific approach to urbanism. In his article *Theory of the Urban Web* he praises the pioneering work of Alexander, the study of urban patterns as fractals conducted by Batty and Longley and the social logic of space promoted by Hillier and Hanson as "notable attempts" to create a "scientific formulation" in architecture and urban design (Salingaros 1998). Salingaros's work is focused on the principle of urban networks that underlies urban form and structure. He states that cities need "mathematical qualities" of connectivity and fractal subdivisions to become living structures (Salingaros 1998).

## 4. Discussions

*Looking backwards* [6]

The traditional city or the city pre - modern urban planning is still evoked as a model to follow and it seems that all the efforts that have been made in understanding cities – deconstructing it, analysing it, categorising it – go in the direction of recovering a distant past. Throughout the history of urban theories there is a chain of thought that cities forms can be manipulated and shaped at our will and by working upon the built environment and its physical features it is possible to achieve better quality of life.

---

[4] Patrick Geddes is considered the first to face the problem of complexity in urban planning according to Batty and Marshall (see Portugali, 2012)

[5] It is a revised and enlarged version of his 1977 book entitled *Fractals: Form, Chance and Dimension*, which in turn was a revised, enlarged, and translated version of his 1975 French book, *Les Objects Fractals: Forme, Hasard et Dimension*.

[6] This is a reference to Edward Bellamy's novel *Looking Backward: 2000-1887*. Describing a utopian urban world this book influenced Ebenezer Howard in his concept of the Garden City (see Hall, 1988).

The distinctions between organic or natural and planned cities mentioned by Batty and Longley (1994) echo the ideas developed by Christopher Alexander on his search for a more naturally designed cities. But Alexander also suggested that a complete understanding of the semi lattice structure that makes natural cities such better places to live in might be beyond our mental capacities as human brains cannot conceive complex set structures in a single mental act (Alexander 1965). This implies that cities could or even should be unplanned by principle (Marshall 2012).

However the paradox of urban planning – designing better future cities without achieving the qualities of the past ones – has not deterred the pursuit of a scientific approach to cities that would eventually lead to a replication of natural urban dynamics. The legacy of theoreticians of the twentieth century remains significant on contemporary urban research and is continually used to support methods and techniques. Researchers as Batty, Salingaros and Hillier have made several references to Alexander and Jacobs. The urban analysis proposed by Kevin Lynch in the 1960s is brought back to present new techniques to measure visual perception with isovists (Morello and Ratti 2009).

***Untamed technologies, untamed cities?***

Systems like GIS has become less expensive to implement, less difficult to learn, more intuitive and above all more customisable. The 1990s brought the IT revolution that popularised the personal computer and the internet and software interfaces became user friendly, giving to the lay public access to powerful computer processing tools. This revolution created a proactive user equipped to question the capabilities and usefulness of software products and also to interact with the IT industry in developing specific applications, many of these with open source software.

Interactive web-based technologies are now on the rise allowing users to explore new ways to share information that is often generated by the users themselves. Another revolutionary wave has appeared after the 2000s with crowd-sourced datasets shared online that are creating a form of GIS without experts, also known as neogeography or geography constructed by users and their cartography do-it-yourself (Warf and Sui 2010) and the Volunteered Geographical Information movement (Goodchild 2007; Elwood et al 2012).

The city built without planners, made of architecture without architects, is not a new idea and throughout the history of urban planning few experiments , although localised and timid in scale, were conducted in that direction (Hall 1988; Cherry 1996). Despite the criticism that it has received, and the frustrations experienced from unsuccessful endeavours, modern urban planning is here to stay.  Releasing cities  from the constrains of planning seems a too scary idea as urban growth is continually challenging governments, organisations and communities all over the world to develop sustainable living in urban environments.